\documentclass[aps,prl,twocolumn,superscriptaddress,nofootinbib]{revtex4}
\pdfoutput=1
\usepackage{hyperref} 

\usepackage[centertags,sumlimits,intlimits,namelimits,reqno]{amsmath}
\usepackage{bbm,amsfonts}
\usepackage{amsthm}

\usepackage{graphicx}
\usepackage{dcolumn}
\usepackage{bm}

\newcommand{\bra}[1]{\langle {#1}|}
\newcommand{\ket}[1]{|{#1}\rangle}

\newcommand{\ketbra}[2]{\ket{#1}\negmedspace\bra{#2}}
\newcommand{\id}{\mathbf{1}}

\def\tr{\mathrm{tr}}

\def\Z{{\mathbbm Z}}
\def\X{{\mathbbm X}}
\def\Y{{\mathbbm Y}}
\def\Id{{\mathbbm 1}}
\newcommand{\mean}[1]{\langle{#1}\rangle}

\begin{document}

\title{Testing contextuality on quantum ensembles with one clean qubit.}

\author{O. Moussa}
 \email{omoussa@iqc.ca}
\author{C. A. Ryan}%
\affiliation{%
Institute for Quantum Computing and 
Department of Physics and Astronomy, 
University of Waterloo, Waterloo, ON, N2L3G1, Canada.}

\author{D. G. Cory}
\affiliation{ Department of Nuclear Science and Engineering, MIT, 
Cambridge, Massachusetts 02139, USA}
\affiliation{
Perimeter Institute for Theoretical Physics, Waterloo, ON, N2J2W9, Canada}

\author{R. Laflamme}%
\affiliation{%
Institute for Quantum Computing and 
Department of Physics and Astronomy, 
University of Waterloo, Waterloo, ON, N2L3G1, Canada.}
\affiliation{
Perimeter Institute for Theoretical Physics, Waterloo, ON, N2J2W9, Canada}

\begin{abstract}
We present a protocol to evaluate the expectation value of the correlations of measurement outcomes for ensembles of quantum systems, and use it to experimentally demonstrate--under an assumption of \emph{fair sampling}--the violation of an inequality that is satisfied by any non-contextual hidden-variables (NCHV) theory. The experiment is performed on an ensemble of molecular nuclear spins in the solid state, using established Nuclear Magnetic Resonance (NMR) techniques for quantum information processing (QIP). 
\end{abstract}
\maketitle

The Bell-Kochen-Specker (BKS) theorem~\cite{Specker60,Bell66,KS67,Mer93a} states that \emph{no} noncontextual hidden-variables (NCHV) theory can reproduce the predictions of quantum mechanics for correlations between measurement outcomes of some sets of observables. Any such set of observables constitutes a proof of the theorem. 
Recently, Cabello~\cite{Cab08} and others~\cite{BBC+08} used BKS proofs to derive a set of inequalities that are satisfied by any NCHV theory but are violated by quantum mechanics for \emph{any} quantum state. 
These inequalities bound certain linear combinations of ensemble averages of correlations between measurement outcomes of compatible observables; thus creating a separation between the predicted outcomes of quantum mechanics, and the bound that is satisfied by NCHV theories. 

This provides an opportunity to test noncontextuality with finite-precision experiments --which has been the subject of contention for many years~\cite{Mey99a, Cab02, Mer99a}-- and without the need for the creation of special quantum states~\cite{SZW+00,HLZ+03,BKS+09}. Already, two experiments, on a pair of trapped $^{40}$Ca$^+$ ions~\cite{KZG+09}, and with single photons~\cite{ARB+09}, have demonstrated this state-independent conflict with noncontextuality. In this letter, we examine testing contextuality on quantum ensembles.

This manuscript is organized as follows. First, we sketch the arguments leading to one of the inequalities derived in~\cite{Cab08}. Then we present an algorithm to estimate the expectation value of the correlations of measurement outcomes for ensembles of quantum systems. 
And lastly, we report and discuss the result of experimentally implementing the algorithm on a 3-qubit ensemble of molecular nuclear spins in the solid state.


\textbf{Inequality -- }For a quantum system prepared according to some state, $\rho$, one can assign simultaneous outcomes $\{\nu(S_k)\}$ of measurements of a set $\{S_k\}$ of \emph{coobservables} (i.e. \emph{co}measureable; mutually \emph{co}mpatible; \emph{co}mmuting).  In this case, the correlation between the measurement outcomes is given by
\begin{equation}
\label{eqcorr}
\pi_{\{S_k\}} = \prod_k \nu(S_k) =  \nu(\prod_k S_k)\,,
\end{equation}
irrespective of the product ordering. Repeating the preparation and measurement many times, and averaging over the outcomes, one obtains an estimate of the ensemble average of the correlation 
$\mean{\pi_{\{S_k\}}}_\rho=\mean{\prod_k \nu(S_k)}_\rho\,.$

For the case where the coobservables $\{S_k\}$ are also dichotomic,  with possible outcomes $\{\nu(S_k)=\pm1\}$, the correlation~\eqref{eqcorr} also takes on the possible values $\pm1$, and the ensemble average satisfies $-1 \le \mean{\pi_{\{S_k\}}}_{\rho} \le +1$. Note, that in this case, these operators are Hermitian and unitary (also known as Quantum Boolean Functions).

Consider any set of observables with possible outcomes $\pm1$ arranged in a $3\times 3$ table such that the observables in each column and each row are coobservable. It has been shown~\cite{Cab08} that, for any NCHV theory,
\begin{equation}
\label{ineq}
\beta = \mean{\pi_{r_1}}+\mean{\pi_{r_2}}+\mean{\pi_{r_3}}
+\mean{\pi_{c_1}}+\mean{\pi_{c_2}}-\mean{\pi_{c_3}} \le 4\;,
\end{equation}
where $\mean{\pi_{r_1}}$ is the ensemble average of the correlation between outcomes of the observables listed in the first row, and so forth. The above inequality is independent of the preparation of the ensemble, provided all terms are estimated for the same preparation. 

\renewcommand{\arraystretch}{1.4}
\begin{table}[h!]
\begin{center}
\begin{tabular}{c|c|c|c|c|c|}
\cline{2-4}\cline{6-6}
& $c_1$
& $c_2$
& $c_3$ 
&& $\prod$ \\
\cline{1-4}\cline{6-6}
 \multicolumn{1}{|c|}{$r_1$}
& $\Z\Id$
& $\Id\Z$
& $\Z\Z$
&& $+\Id$\\
\cline{1-4}\cline{6-6}
 \multicolumn{1}{|c|}{$r_2$}
& $\Id\X$
& $\X\Id$
& $\X\X$
&& $+\Id$\\
\cline{1-4}\cline{6-6}
 \multicolumn{1}{|c|}{$r_3$}
& $\Z\X$
& $\X\Z$
& $\Y\Y$
&& $+\Id$\\
\cline{1-4}\cline{6-6}
\multicolumn{6}{c}{}\\ [-11pt]
\cline{1-4}
 \multicolumn{1}{|c|}{$\prod$}
& $+\Id$
& $+\Id$
&$-\Id$\\
\cline{1-4}
\end{tabular}
\caption{\label{tabobs} List of the 2-qubit observables used to show that quantum mechanics violates inequality~\eqref{ineq}. This list has been used by Peres~\cite{Per90a} and Mermin~\cite{Mer90a} as a BKS proof for 4 dimensional systems. $\{\Id,\X,\Z,\Y\}$ are the single qubit pauli operators, and, e.g. $\Z\X:=\Z\otimes\X$ indicates a measurement of the pauli-$\Z$ on the first qubit and pauli-$\X$ on the second. 
}
\end{center}
\end{table}%

Now, consider a 2-qubit system (e.g. 2 spin-$\tfrac{1}{2}$ particles), and the set of observables listed in Table~\ref{tabobs}. For any NCHV theory, the inequality~\eqref{ineq} holds for the correlations between measurement outcomes of the coobservables listed in each row and column, where, for e.g., $\mean{\pi_{r_1}}=\mean{\pi_{\{\Z\Id,\Id\Z,\Z\Z\}}}=\mean{\Z\Id\cdot\Id\Z\cdot\Z\Z}$, and so forth. 

On the other hand, according to quantum mechanics, the ensemble average $\mean{\pi_{\{S_k\}}}_{\rho}$ is given by $\tr(\rho\ \prod_k S_k)$. Thus, for a set of coobservables whose product is proportional to the unit operator --as is the case for all rows and columns of Table~\ref{tabobs}-- the quantum mechanical prediction of the ensemble average of the correlation is equal to the proportionality constant, independent of the initial preparation of the system. Hence, the quantum mechanical prediction for $\beta$ is 6, which violates inequality~\eqref{ineq}. 

\textbf{Algorithm -- } To measure the correlation between a set of coobservables, consider introducing an ancillary (probe) qubit, and applying a transformation $U_{S_k}$ to the composite system for each observable $S_k$, in a manner reminiscent of coherent syndrome measurement in quantum error correction~\cite{DS96}. 
For an observable $S$ with the spectral decomposition $S= P_+-P_-$, where $P_+$ and $P_-$ are the projectors on the $+1$ and $-1$ eigenspaces of $S$, the transformation $U_S$ is defined as $U_S=\id_2 \otimes P_++\Z \otimes P_-\,$. That is to say, if the system is in a -1 eigenstate of $S$, apply a phase flip (pauli-$\Z$) to the probe qubit, and if it is in a +1 eigenstate, do nothing. This transformation can also be expressed as a controlled operation dependent on the state of the probe qubit:
\begin{equation*}
\begin{split}
U_S
&=
\id_2 \otimes P_+
+\Z \otimes P_-\\
&=
\tfrac{1}{2}\left(\id_2+\Z\right) \otimes (P_+ + P_-)
+ \tfrac{1}{2} \left(\id_2-\Z\right) \otimes (P_+ - P_-)\\
&=
\ketbra{0}{0} \otimes \id_d
+ \ketbra{1}{1} \otimes S\,,
\end{split}
\end{equation*}
which is unitary for $S$ unitary. If the system is initially prepared according to $\rho$, and the probe qubit in the $+1$ eigenstate of the pauli-$\X$ operator, $\ket{+}=(\ket{0}+\ket{1})/\sqrt2$, the possible outcomes of pauli-$\X$ measurement on the probe qubit is $\pm 1$, with probabilities $p(\pm1)$ given by:
\begin{equation*}
\begin{split}
p(\pm1)
=&
\tr_{a+s} \left[ U_S (\ketbra{+}{+} \otimes \rho) U_S^\dagger  \; (\ketbra{\pm}{\pm} \otimes \id_d) \right]\\
=&
\tr_{s} \left[ \bra{\pm} (\id_2 \otimes P_+ +\Z \otimes P_-) (\ketbra{+}{+} \otimes \rho) (...)^\dagger \right\ket{\pm}]\\
=&
\tr_s \left[ P_{\pm} \rho\  \right]\,, \\
\end{split}
\end{equation*}
and the ensemble average
\begin{equation}
\begin{split}
\mean{\X\otimes \Id_d}
&= 
+p(+1)-p(-1)\\
&=
\tr_s \left[ P_{+}\, \rho\,  \right] - \tr_s \left[ P_{-}\, \rho\,\right]  \\
&=
\mean{S}_\rho\,.
\end{split}
\end{equation}
\begin{figure}
\begin{center}
\includegraphics[width=2.4in]{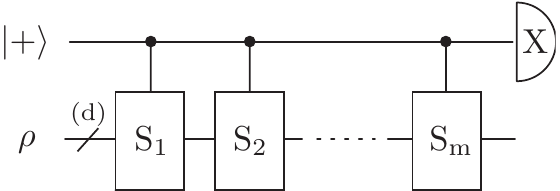}
\caption{\label{figCirc}  A quantum network to encode the correlation between the outcomes of measurements $\{S_k\}_{k=1...m}$ on a $d$-dimensional system, in the phase of a probe qubit state. Repeating this procedure for the same preparation $\rho$ and averaging the outcome of the measurement on the probe qubit gives the ensemble average $\mean{S_1S_2 \cdots S_m}_\rho$. Alternatively, for an ensemble of quantum systems initially prepared according to $\rho$, on which operations are applied in parallel to the individual systems, an ensemble measurement readily produces $\mean{S_1S_2\cdots S_m}_\rho$. }
\end{center}
\end{figure}
Thus, to measure the ensemble average of the correlation between a set of coobservables, one prepares a probe qubit in the +1 eigenstate of $\X$, and the system according to $\rho$. As shown in Figure~\ref{figCirc}, one then applies the unitaries $S_k$ in succession to the system, controlled on the state of the probe qubit. Since, by definition, all $S_k$ mutually commute, then the order of their application has no bearing on the measurement outcome.  Repeating this procedure, and averaging the outcome of the measurement on the probe system produces the correlation between this set of observables. 
Alternatively, one could prepare an ensemble of systems according to $\rho$; apply the transformations $U_S$ in parallel to each member of the ensemble; and perform a bulk ensemble measurement to estimate $\mean{\pi_{\{S_k\}}}_{\rho}$. This alleviates the need for isolation of single quantum systems, and the repeated application of single shot, projective measurement. 

Since inequality~\eqref{ineq} is valid for any preparation $\rho$, then one is free to choose to prepare the system according to the maximally mixed state. In which case, only one qubit --the probe system-- is not maximally mixed. This corresponds to the model of computation known as Deterministic Quantum Computation with one clean qubit (DQC1)~\cite{KL98a}. 

\textbf{Two models --}
Suppose the measurement process on the probe qubit was $\epsilon$-efficient, i.e. returning a faithful answer $\epsilon$ fraction of the time, and otherwise a uniformly distributed random outcome. The probabilities $p(\pm1)$ of obtaining outcomes $\pm1$ will be modified to: $p(\pm1)= \tfrac{1-\epsilon}{2} + \epsilon\ \tr_s \left[ P_{\pm} \rho\  \right]$, and the ensemble average to $\mean{\X\otimes \Id_d}= \epsilon\ \mean{S}_\rho$. One can then estimate the expectation value $\mean{S}_\rho$ under an assumption of fair sampling and knowing the value of $\epsilon$, which can be established from $\epsilon\mean{\Id}_\rho$.
This model is equivalent to one where the probe system is initially in the mixed state $(1-\epsilon)\tfrac{\Id_2}{2}+\epsilon \ketbra{+}{+}$, provided the reduced dynamics on the probe qubit from preparation to measurement is represented by a unital map; i.e. a map that preserves the totally mixed state. To see this, suppose we prepare the probe qubit in some state $\rho_a$ and then apply some transformation to the composite system, whose reduced dynamics on the probe qubit is described by a unital linear map $\Lambda$. An $\epsilon$-efficient measurement of $\X=\ketbra{+}{+}-\ketbra{-}{-}$ has two possible outcomes $\pm1$ with probabilities
\begin{equation*}
\begin{split}
p(\pm1)
&=
\tfrac{1-\epsilon}{2} + \epsilon\ \tr \left[\ \ketbra{\pm}{\pm}\ \Lambda(\rho_a)\  \right]\\
&=
\tfrac{1-\epsilon}{2}\  \tr \left[\ \ketbra{\pm}{\pm}\  \right]+ \epsilon\ \tr \left[\ \ketbra{\pm}{\pm}\ \Lambda(\rho_a)\  \right]\\
&=
\tfrac{1-\epsilon}{2}\  \tr \left[\ \ketbra{\pm}{\pm}\  \Lambda(\Id_2)\ \right]+ \epsilon\ \tr \left[\ \ketbra{\pm}{\pm}\ \Lambda(\rho_a)\  \right]\\
&=
\tr \left[\ \ketbra{\pm}{\pm}\  \Lambda\left((1-\epsilon)\tfrac{\Id_2}{2}+ \epsilon\ \rho_a\right)\  \right],
\end{split}
\end{equation*}
which are precisely the statistics one obtains in case the probe qubit is initially in the state $(1-\epsilon)\tfrac{\Id_2}{2}+ \epsilon\ \rho_a$, and the measurement process is faithful. 

\begin{figure}
\begin{center}
\begin{tabular}{ll}
(r$_1$)&(c$_1$)\\
\includegraphics[width=1.5in]{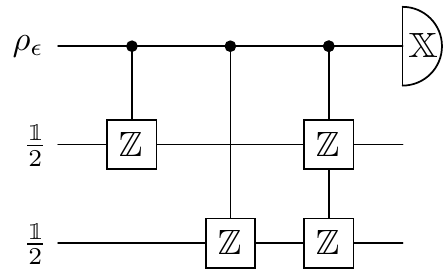}&
\includegraphics[width=1.5in]{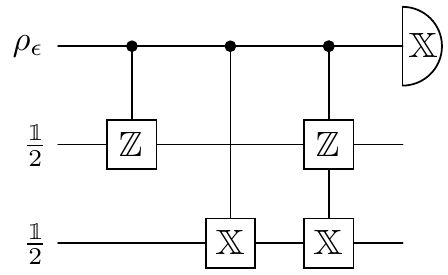}\\
(r$_2$)&(c$_2$)\\
\includegraphics[width=1.5in]{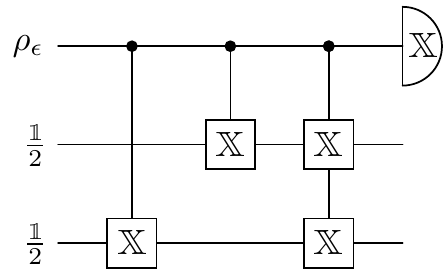}&
\includegraphics[width=1.5in]{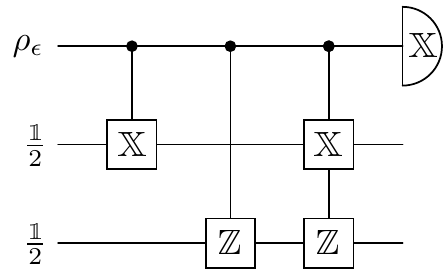}\\
(r$_3$)&(c$_3$)\\
\includegraphics[width=1.5in]{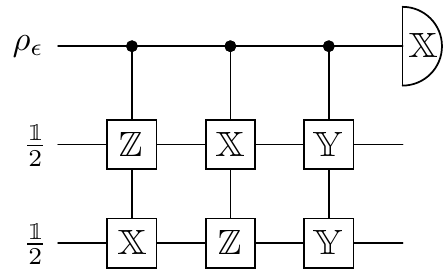}&
\includegraphics[width=1.5in]{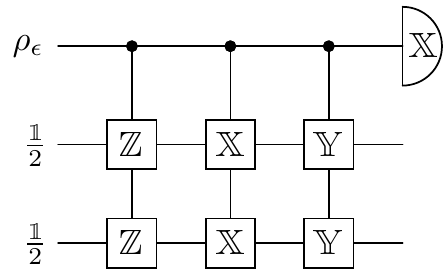}
\end{tabular}
\caption{\label{figcircuits} The quantum networks for the six experiments to estimate $\beta$ as given in~\eqref{ineq}. The ensemble is initially prepared according to $\rho_\epsilon\otimes\tfrac{\Id_2}{2}\otimes\tfrac{\Id_2}{2}$, where $\rho_\epsilon= (1-\epsilon)\tfrac{\Id_2}{2}+ \epsilon \ketbra{+}{+}$, and $\tfrac{\Id_2}{2}$ is the single-qubit maximally mixed state. }
\end{center}
\end{figure}

\textbf{Experiment --} 
We implement the algorithm described above to perform an experimental measurement of the correlations as described in inequality~\eqref{ineq} on an ensemble of nuclear spins in the solid state using established NMR techniques for QIP~\cite{BMR+06,RMB+08}. Figure~\ref{figcircuits} shows the six experiments required to estimate the six terms in~\eqref{ineq}. The pulse sequence implementing the measurement of some observable is the same whether it is being measured with the coobservables listed in its row or column.
%
%

\begin{figure}
\hspace{-.4in}
\begin{minipage}{0.4\linewidth}
\includegraphics[scale=.18]{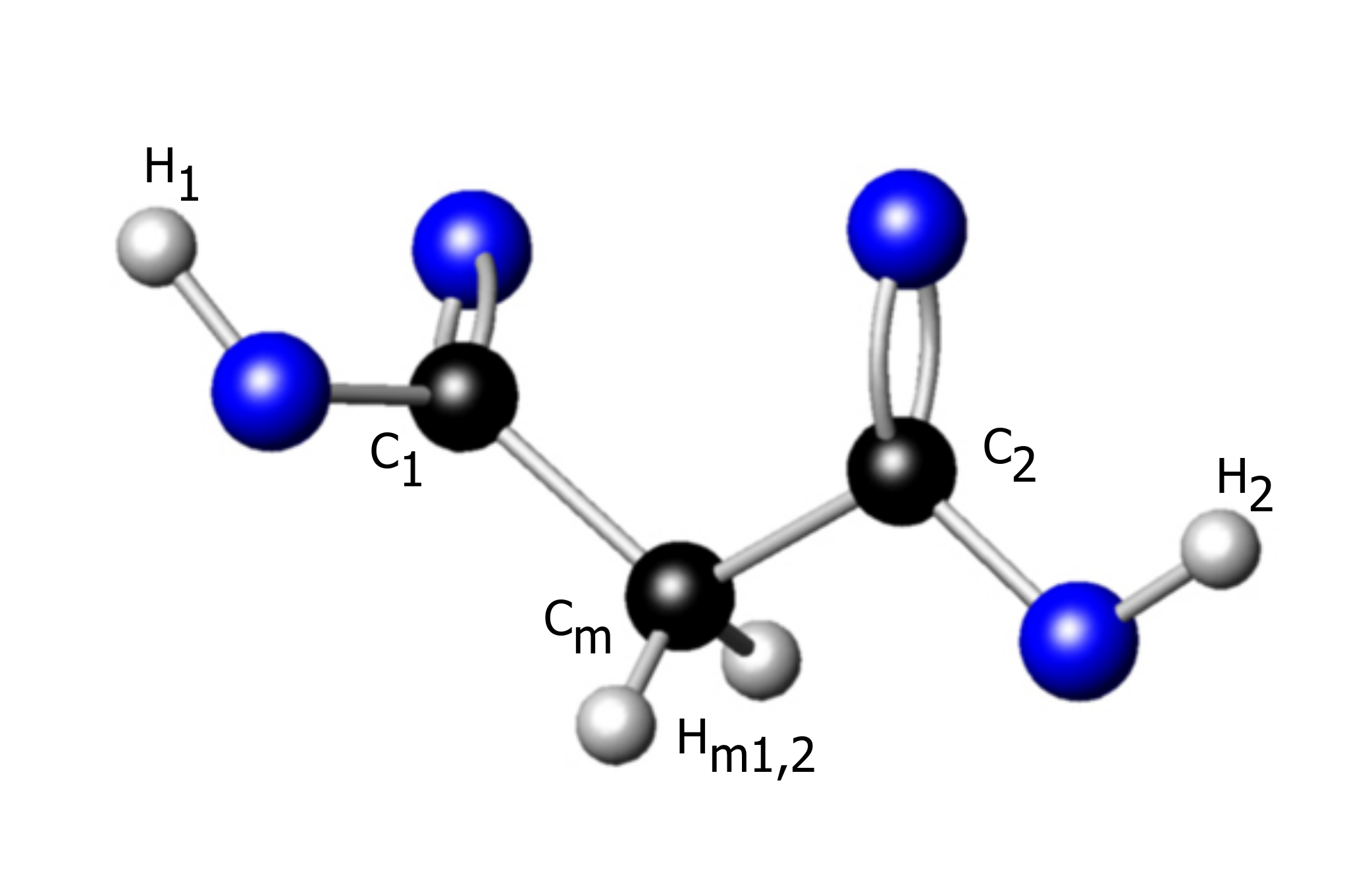}
\end{minipage}
\hspace{0.5cm}
\begin{minipage}{0.35\linewidth}
 \begin{small}
\begin{tabular}{|c|c|c|c|}
\hline
kHz&C$_1$&C$_2$&C$_m$\\
\hline
C$_1$&6.380&0.297&0.780\\
\hline
C$_2$&-0.025&-1.533&1.050\\
\hline
C$_m$&0.071&0.042&-5.650\\
\hline
\end{tabular}
\end{small}
\end{minipage}\\
\includegraphics[width=3.2in]{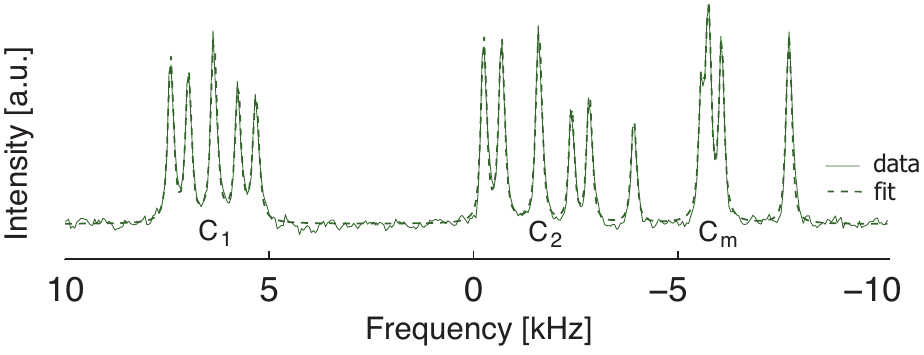}
\caption{\label{figmalonic} Malonic acid (C$_3$H$_4$O$_4$) molecule and Hamiltonian parameters (all values in kHz).  Elements along the diagonal represent chemical shifts, $\omega_i$, with respect to the transmitter frequency (with the Hamiltonian $\sum_i \pi \omega_i \Z_i$). Above the diagonal are dipolar coupling constants ($\sum_{i<j} \pi D_{i,j} (2\ \Z_i\Z_j - \X_i\X_j - \Y_i\Y_j$), and below the diagonal are J coupling constants, ($\sum_{i<j} \tfrac{\pi}{2} J_{i,j} (\Z_i\Z_j + \X_i\X_j + \Y_i\Y_j$).  An accurate natural Hamiltonian is necessary for high fidelity control and is obtained from precise spectral fitting of (also shown) a proton-decoupled $^{13}$C spectrum following polarization-transfer from the abundant protons. The central peak in each quintuplet is due to natural abundance $^{13}$C nuclei present in the crystal at $\sim 1\%$. (for more details see~\cite{BMR+06,RMB+08} and references therein.) }
\vspace{.1in}
\end{figure}

The experiments were performed in a static field of $7.1$T using a purpose-built probe. The sample is a macroscopic single crystal of Malonic Acid (C$_3$H$_4$O$_4$), where a small fraction ($\sim 3\%$) of the molecules are triply labeled with $^{13}$C to form an ensemble of processor molecules. During computation, these processors are decoupled from the 100\% abundant protons in the crystal by applying a decoupling pulse sequence~\cite{FKE00} to the protons. Shown in Figure~\ref{figmalonic} is a proton-decoupled $^{13}$C spectrum, following polarization-transfer from the abundant protons, for the particular orientation of the crystal used in this experiment. A precise spectral fit gives the Hamiltonian parameters (listed in the inset table in Figure~\ref{figmalonic}), as well as the free-induction dephasing times, $T_2^*$, for the various transitions; these average at $\sim2$ms. The dominant contribution~\cite{BMR+06} to $T_2^*$ is Zeeman-shift dispersion, which is largely refocused by the control pulses. Other contributions are from intermolecular $^{13}$C-$^{13}$C dipolar coupling and, particularly for C$_m$, residual interaction with neighboring protons. 

The carbon control pulses are numerically optimized to implement the required unitary gates using the GRAPE~\cite{KRK+05} algorithm. Each pulse is $1.5$ms long, and is designed~\cite{RNL+08} to have an average Hilbert-Shmidt fidelity of $99.8\%$ over appropriate distributions of Zeeman-shift dispersion and control-fields inhomogeneity.

\begin{figure}
\begin{center}
\includegraphics[width=3.2in]{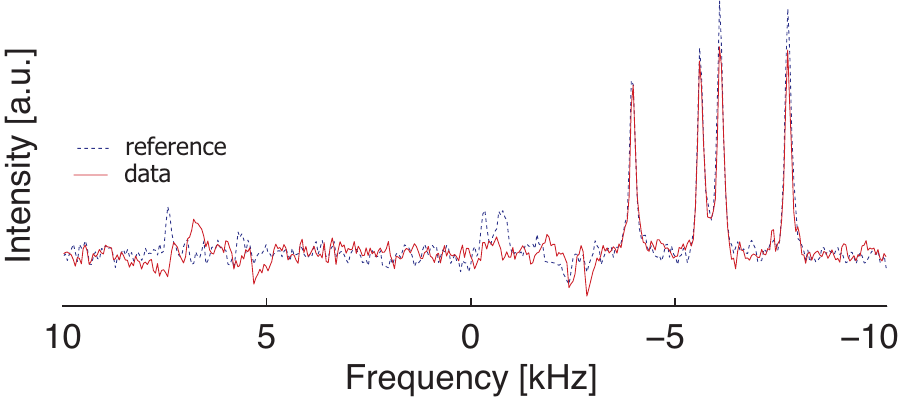}
\caption{\label{figresults}(color online) Summary of experimental results. Shown are (in dashed blue) the spectrum produced by the initial preparation procedure, $\rho_\epsilon\otimes\tfrac{\Id_2}{2}\otimes\tfrac{\Id_2}{2}$, establishing a reference for $\epsilon$; and (in solid red) the sum of the six spectra corresponding to the six terms in $\beta$ of inequality~\eqref{ineq} with the appropriate signs, scaled by $\tfrac{1}{6}$ to compare with the reference. } 
\end{center}
\end{figure}

The two spin-$\tfrac{1}{2}$ nuclei C$_1$ and C$_2$, constituting the system on which the measurements are performed, are initially prepared according to the totally mixed state. C$_m$, representing the probe qubit, is initially prepared according to $\rho_\epsilon= (1-\epsilon)\tfrac{\Id_2}{2}+ \epsilon \ketbra{+}{+}$, where $\ket{+}=\tfrac{1}{\sqrt2}(\ket{0}+\ket{1})$, and the computational basis, $\{\ket{0},\ket{1}\}$, is the eigenbasis of the single spin Zeeman Hamiltonian.
Figure~\ref{figresults} shows the spectrum observed for the initial preparation, as well as the average of the spectra from the six experiments, representing the six terms in $\beta$ of inequality~\eqref{ineq} with the appropriate signs. Fitting the observable spectra, taking into consideration the effects of strong coupling, we estimate the value of $\beta$ to be $5.2\pm0.1$, in violation of inequality~\eqref{ineq}. The uncertainty on $\beta$ is propagated from the goodness-of-fit figure of merit ascribed to the spectral fitting process.

\begin{figure}
\begin{center}
\includegraphics[width=3.3in]{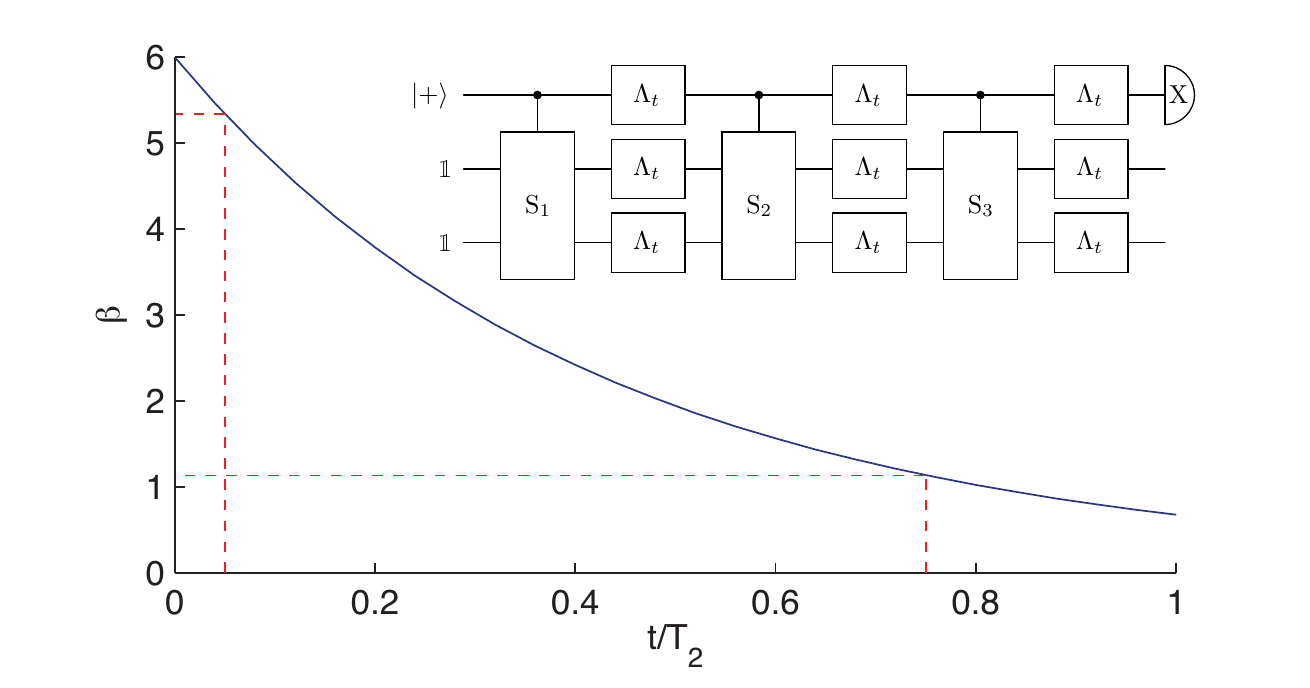}
\caption{\label{figdecoherence}  Numerical simulation results of a simple model of decoherence (inset) showing the expected variation of $\beta$ as a function of the ratio of the pulse-length, $t$, to an effective dephasing time, $T_2$. The dashed lines indicate bounds on the expected performance of the current experiment; for pulse length of $1.5$ms, and effective decoherence times of $2$ms ($\sim T_2^*$) and $30$ms ($\sim$ intrinsic coherence times~\cite{BMR+06}), the value of $\beta$ is expected to be $1.1$ and $5.3$ respectively. } 
\vspace{-.3in}
\end{center}
\end{figure}

Decoherence, as it is wont to do, causes deviations from the idealized closed-system dynamics. To examine its effect, we numerically simulate the dynamics of a simple model (shown in Figure~\ref{figdecoherence}) in which each ideal transformation is followed by a symmetric error of a three-fold tensor product of a single-qubit dephasing map,  $\Lambda(\rho)$, given by the operator sum representation
$  \rho \rightarrow \Lambda(\rho) = \sum_\kappa A_\kappa \rho A_\kappa^\dag $, where  $A_0=\left(\begin{smallmatrix}
1 & 0 \\
0 & \sqrt{1-\eta}\\
\end{smallmatrix}\right),\,
 A_1=\left( \begin{smallmatrix}
1 & 0 \\
0 & \sqrt{\eta}\\
\end{smallmatrix}\right)\,,
$
and the parameter $\eta=1-\exp{(-t/T_2)}$ depends on the ratio of the pulse-length, $t$, to an effective dephasing time, $T_2$. Using appropriate estimates~\cite{BMR+06} of this dephasing time, one is able to largely explain the deviation of the experimental result from the prediction of quantum mechanics in ideal conditions.

\textbf{Conclusion --}
We have presented a protocol to directly measure correlations between measurement outcomes,  utilizing an ancillary (probe) two dimensional system, with the purpose of testing quantum contextuality. Conveniently, it can be used directly on ensembles of quantum systems, without the need for repeated projective measurement on single systems. Additionally, it can be straightforwardly extended to test similar inequalities on higher-dimensional systems. 
Our experimental results demonstrate
--under the assumption of fair sampling--
that a three-qubit deterministic quantum computer with one clean qubit reveals correlations between measurement outcomes that cannot be explained by any NCHV theory. 

\textbf{Acknowledgements --} 
This work benefitted from discussions with J. Emerson, A. Cabello, M. Laforest, and J. Baugh. This research was supported by NSERC, CIFAR, and the Premier Discovery Awards. This work was supported in part by the National Security Agency (NSA) under Army Research Office (ARO) contract number W911NF-05-1-0469.

\end{document}